\documentclass[11pt,twoside]{article}

\usepackage{asp2006}
\usepackage{graphicx}
\usepackage{lscape}

\begin{document}
\title{Calibration of Star Formation Rate Tracers with Population Synthesis 
Models: the \boldmath $L_{\rm X}$ to $L_{\rm FIR}$ Ratio}   
\author{H\'ector Ot\'i--Floranes$^1$ and Jose Miguel Mas--Hesse$^2$}  
\affil{$^1$LAEFF--INTA, POB 78, 28691 Villanueva de la Ca\~nada, Spain \\ 
$^2$Centro de Astrobiolog\'{\i}a (CSIC--INTA), 28850 Torrej\'on de Ardoz, Spain}

\begin{abstract}
The study of massive star formation needs some basic tools, among which reliable
SFR tracers are crucial ones. We are presently revising the calibration of SFR
tracers at different wavelength ranges using last generation evolutionary
population synthesis codes. The FIR luminosity produced by the heated dust in
star forming regions is commonly used to characterize a starburst. The X-ray
luminosity has been found to show a narrow correlation with it, and therefore it
has been proposed as an additional tracer of potential interest for high $z$
galaxies, whose FIR emission is redshifted to the sub-mm range. In this
communication  we analyze the evolution of the X-ray to FIR luminosities ratio
as predicted by the CMHK population synthesis models, both for constant stellar
formation and instantaneous bursts. The results are compared to some sample data
taken from the literature. The conclusion drawn from the comparison is that the empirical calibration of the soft X-ray luminosity seems a valid SFR tracer only for starbursts around a rather short period of time.
\end{abstract}

\section{The \boldmath $L_{\rm X}$/$L_{\rm FIR}$ ratio}

We have computed the X-ray and far infrared (FIR) luminosities expected in a
massive, young starburst using the CMHK (Cervi\~{n}o, Mas-Hesse \& Kunth)
synthesis population models \citep{Oti_Cer94,Oti_Cer02}. A Salpeter Initial Mass Function ($\phi(m)\sim m^{-2.35}$) has been assumed, with masses within the
range $2 \mbox{--} 120 \, M_{\sun}$, 2 different star formation regimes --
instantaneous burst (IB) or constant star formation rate (CSFR) -- and solar
metallicities. We have analyzed the first 30 Myrs after the onset of a massive
starburst episode.    

The X-ray luminosity is calculated in the range $0.4 \mbox{--} 2.4$ keV. The
models compute only the emisison by the diffuse gas heated by the mechanical
energy released by stellar winds and supernova remnants. The X-ray emission is
mo\-deled by Raymond-Smith thermal plasmas with a range of temperatures typical of
starburst galaxies (between 0.5 and 1.0 keV). The fraction of mechanical ener\-gy
$\epsilon_{\rm eff}$ effectively heating the gas up to X-rays temperatures is
left as a free parameter. The contribution by X-ray binaries and stellar
atmospheres has not been considered, since it should be negligible (Cervi\~{n}o et al. 2002).  

A thermal equilibrium of dust has been assumed, which implies all energy
absorbed by dust being reemitted in the far infrared range. The models assume
that a fraction $(1-f)$ of Lyman continuum photons are directly absorbed by the
dust. The absorption of of UV--optical continuum photons is parameterized
through $E(\bv)$. The values considered  were $f=0.3$ and $E(\bv)=1$.

\begin{figure*}[!ht]
\begin{center}
\includegraphics[height=3.94 cm,bb=-1 37 715 521 dpi,clip=true]{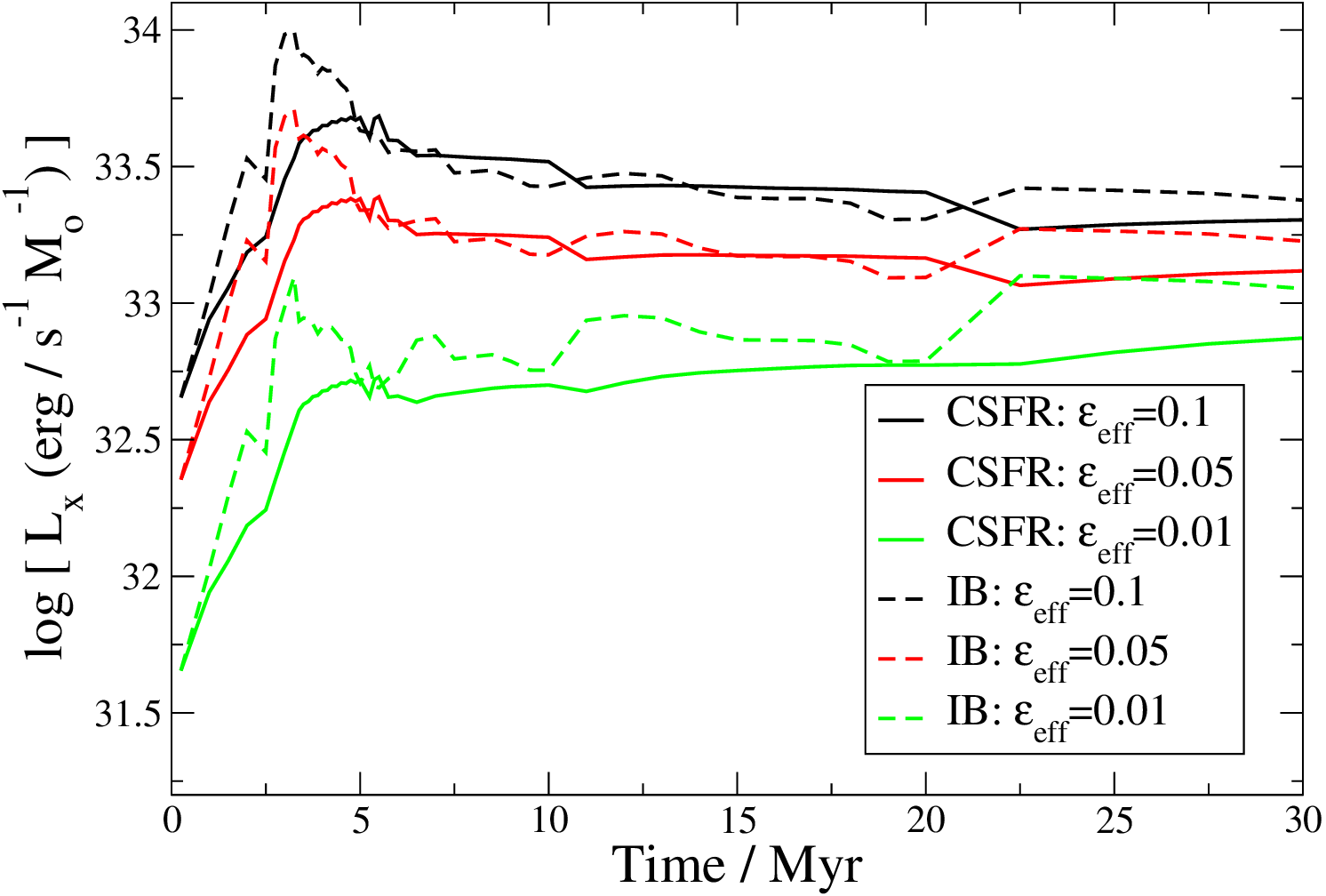}
\hspace{0.75 cm}
\includegraphics[height=3.94 cm,bb=29 37 715 528 dpi,clip=true]{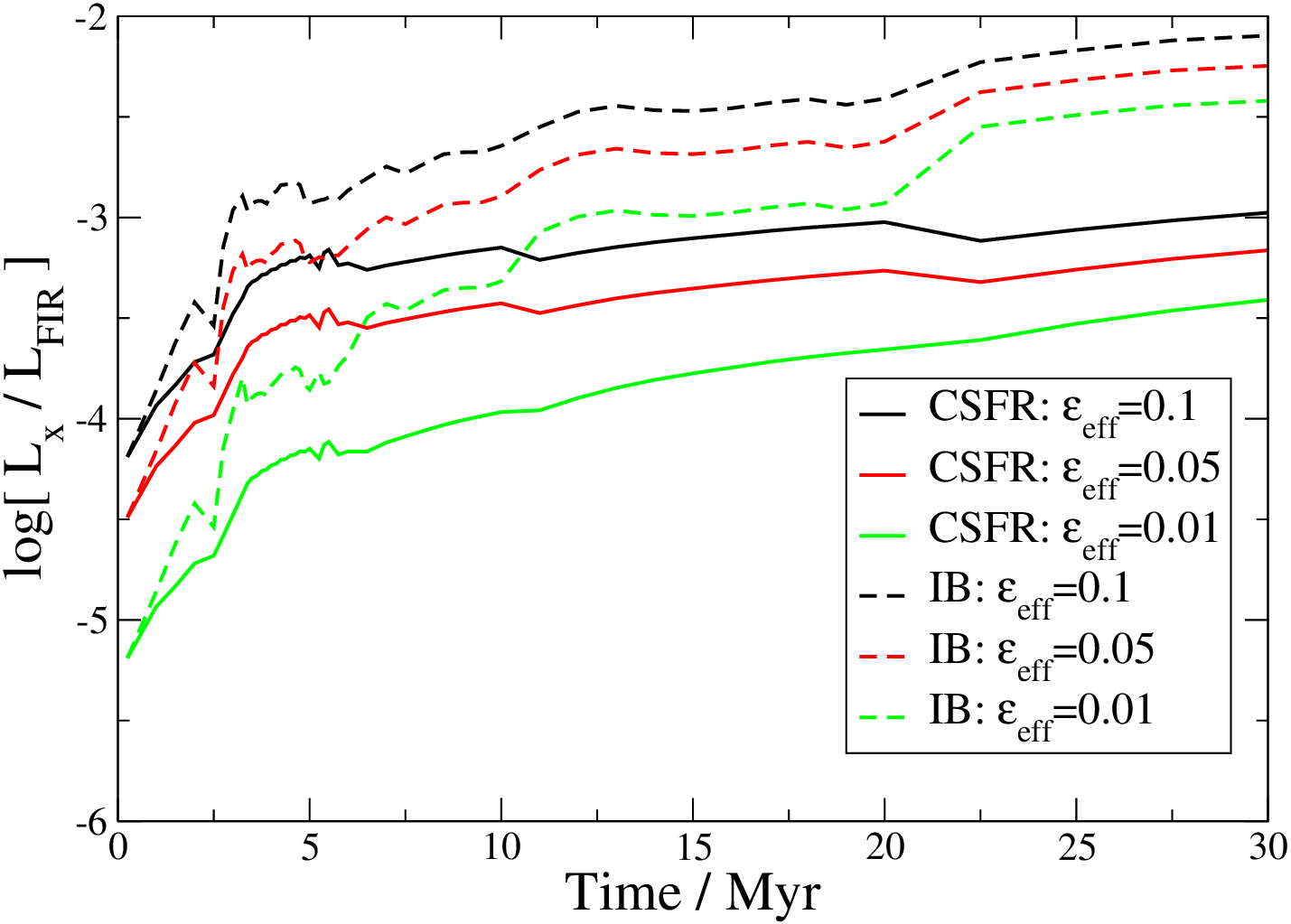}\\
\vspace{0.3 cm}
\hspace{0.75 cm}
\includegraphics[width=5.5 cm,bb=48 31 789 528 dpi,clip=true]{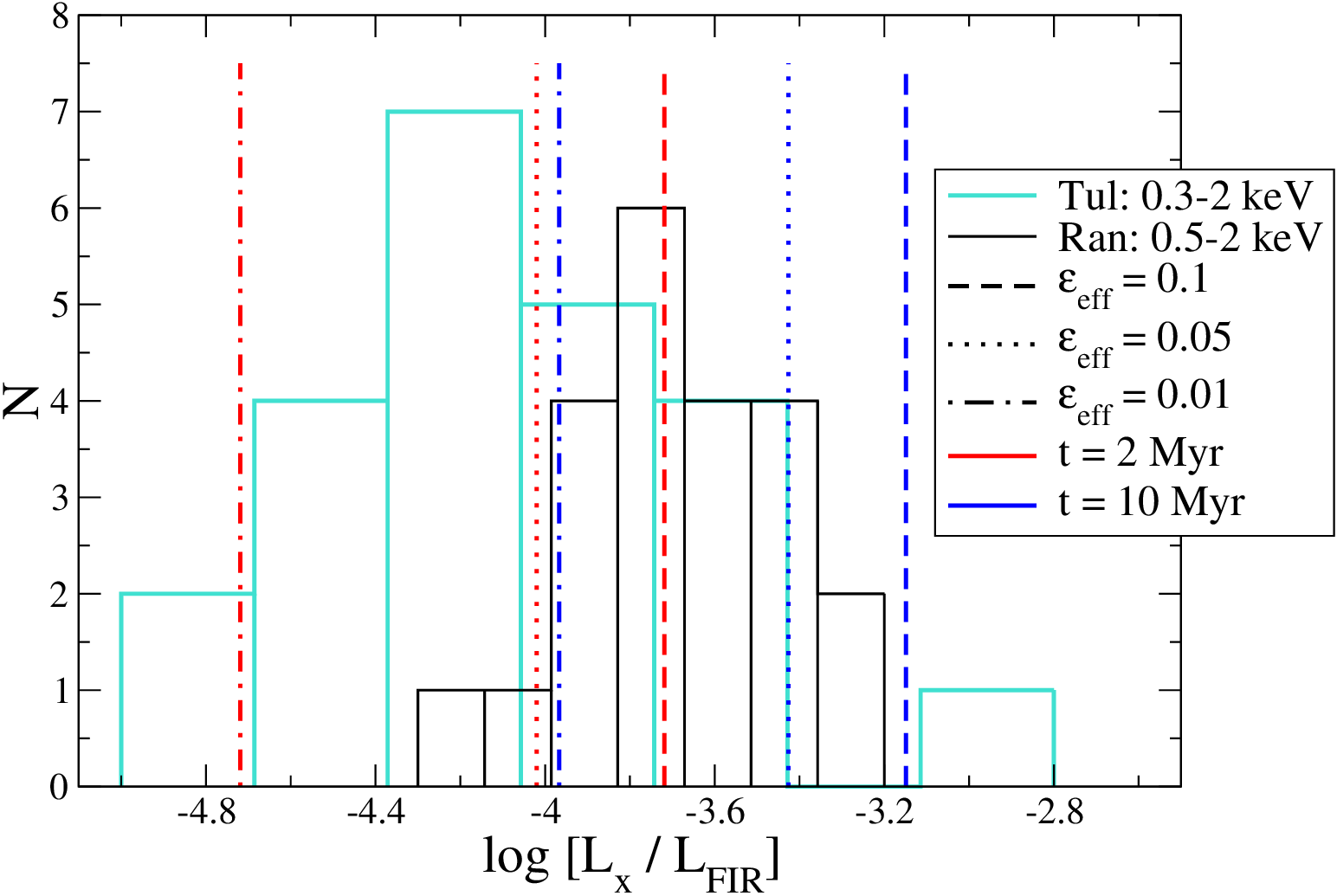}
\hspace{0.75 cm}
\includegraphics[width=5.5 cm,bb=48 31 787 528 dpi,clip=true]{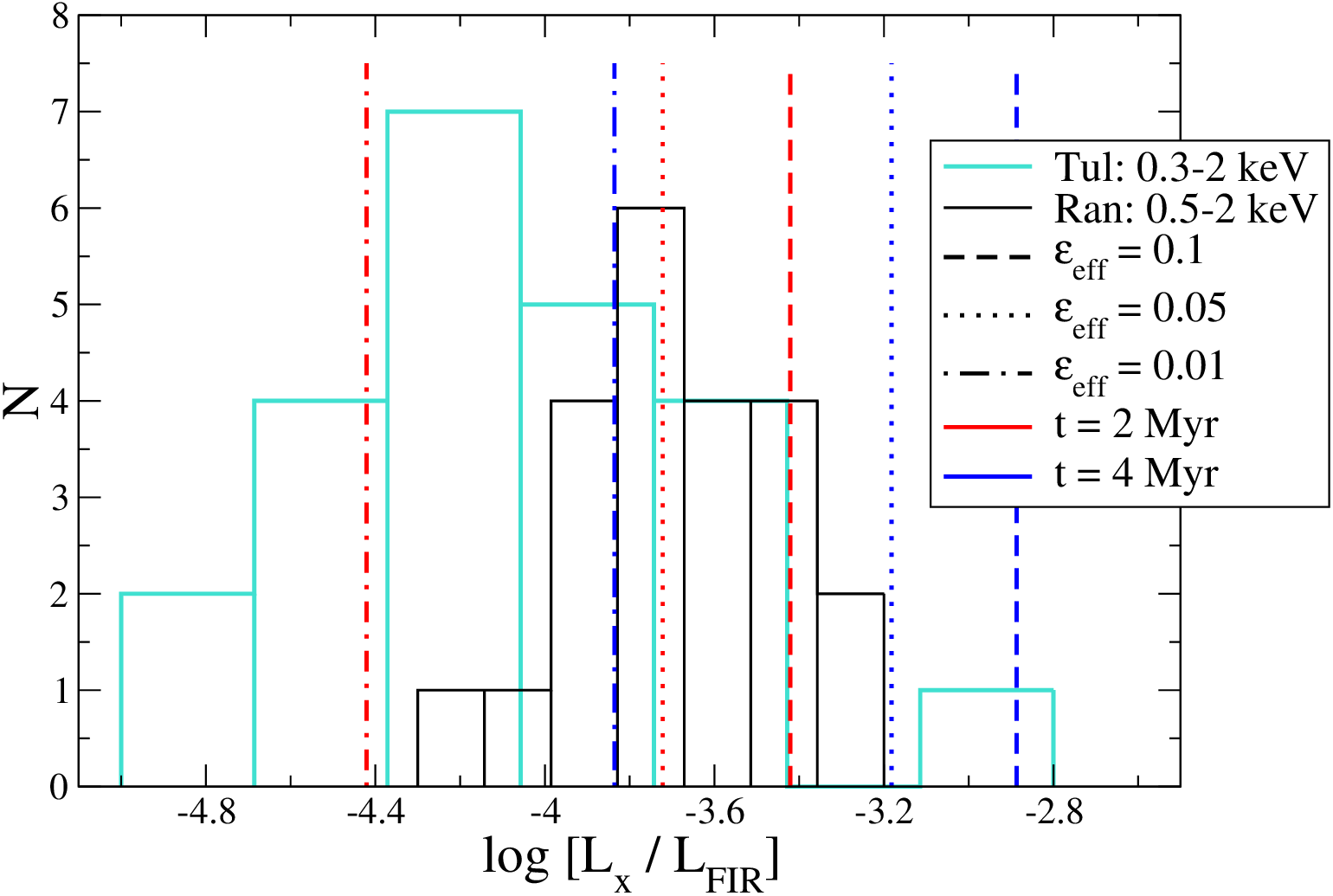}
\end{center}
\caption{Evolution of $L_{\rm X}$ (top left) and $L_{\rm X}$ to $L_{\rm FIR}$ ratio 
(top right) predicted by the synthesis models. 
Histograms of the  $L_{\rm X}$
to $L_{\rm FIR}$ ratio for the samples \emph{Ran} and \emph{Tul} considered,
together with model predicitions: CSFR (bottom left) and  IB (bottom right) cases.}
\label{fig:figu1}
\end{figure*}

Figure~\ref{fig:figu1} shows the evolution of the soft X--ray luminosity and the
\boldmath $L_{\rm X}$/$L_{\rm FIR}$ ratio, as predicted by the models assuming
different star formation regimes and efficiencies. We want to stress that this
ratio is strongly dependent on the evolutionary state of the starburst, spanning
around 2 orders of magnitude within few million years.

The data samples from \citet{Oti_Ran03} and \citet{Oti_Tul06} (hereinafter,
\emph{Ran} and \emph{Tul} respectively) were chosen in order to compare with the
predictions from the models. As shown in the bottom panels of Figure~\ref{fig:figu1}, the
observational ratios can be well reproduced by the synthesis models considering
efficiency values in the range $\epsilon_{\rm eff}=0.01 \mbox{--}0.1$, typical
of starburst galaxies (Cervi\~{n}o et al. 2002).  Moreover, our results show that the
dispersion observed in the samples ratios can be explained as being mostly an  evolutionary
effect.  

Therefore we conclude that the empirical calibration of the soft X-ray luminosity as an SFR (star formation rate) tracer has to be taken with care, since it would be valid only for starbursts around a relatively short period of time.

\end{document}